\documentclass{sigchi}

\CopyrightYear{2020}
\setcopyright{acmcopyright}
\doi{10.1145/3313831.3376428}
\isbn{978-1-4503-6708-0/20/04}
\conferenceinfo{CHI'20,}{April  25--30, 2020, Honolulu, HI, USA}

\usepackage{balance}       
\usepackage{graphics}      
\usepackage[T1]{fontenc}   
\usepackage{txfonts}
\usepackage{mathptmx}
\usepackage[pdflang={en-US},pdftex]{hyperref}
\usepackage{color}
\usepackage{booktabs}
\usepackage{textcomp}

\usepackage{microtype}        
\usepackage{ccicons}          

\usepackage{todonotes}

\usepackage{tabularx}
\usepackage{graphicx}
\usepackage{adjustbox}
\usepackage{color,soul}

\usepackage{xspace}
\usepackage{multirow, booktabs}
\usepackage{makecell}
\usepackage{graphicx}
\usepackage{subcaption}
\usepackage[utf8]{inputenc}

\newcommand*{\eg}{\textit{e.g.,}\@\xspace}
\newcommand*{\ie}{\textit{i.e.,}\@\xspace}
\newcommand{\etal}{\textit{et al}.\@}

\def\plaintitle{Crowdsourcing the Perception of Machine Teaching}

\def\emptyauthor{}
\def\plainkeywords{teachable interfaces; 
interactive machine learning; object recognition; crowdsourcing; personalization}

\makeatletter
\def\url@leostyle{%
  \@ifundefined{selectfont}{
    \def\UrlFont{\sf}
  }{
    \def\UrlFont{\small\bf\ttfamily}
  }}
\makeatother
\def\pprw{8.5in}
\def\pprh{11in}

\definecolor{linkColor}{RGB}{6,125,233}
\hypersetup{%
  pdftitle={\plaintitle},
  pdfauthor={\emptyauthor},
  pdfkeywords={\plainkeywords},
  pdfdisplaydoctitle=true, 
  bookmarksnumbered,
  pdfstartview={FitH},
  colorlinks,
  citecolor=black,
  filecolor=black,
  linkcolor=black,
  urlcolor=linkColor,
  breaklinks=true,
  hypertexnames=false
}

\toappear{\scriptsize Permission to make digital or hard copies of all or part of this work for personal or classroom use is granted without fee provided that copies are not made or distributed for profit or commercial advantage and that copies bear this notice and the full citation on the first page. Copyrights for components of this work owned by others than ACM must be honored. Abstracting with credit is permitted. To copy otherwise, or republish, to post on servers or to redistribute to lists, requires prior specific permission and/or a fee. Request permissions from permissions@acm.org. \\
{\emph{CHI '20, April 25--30, 2020, Honolulu, HI, USA.} } \\
\copyright~2020 Association for Computing Machinery. \\
ACM ISBN 978-1-4503-6708-0/20/04\ ...\$15.00. \\
http://dx.doi.org/10.1145/3313831.3376428 }

\begin{document}

\title{\plaintitle}

\numberofauthors{1}
\author{%
    \alignauthor{Jonggi Hong\textsuperscript{1}, Kyungjun Lee\textsuperscript{1}, June Xu\textsuperscript{2}, Hernisa Kacorri\textsuperscript{3,1}\\
        \affaddr{\textsuperscript{1}Computer Science, \textsuperscript{2}Electrical and Computer Engineering, and \textsuperscript{3}Information Studies}\\ 
        \affaddr{University of Maryland, College Park, MD, USA}\\
        \email{jhong12@umd.edu, kjlee@cs.umd.edu, junexu@terpmail.umd.edu, hernisa@umd.edu}}\\
}

\maketitle

\begin{abstract}
  Teachable interfaces can empower end-users to attune machine learning systems to their idiosyncratic characteristics and environment by explicitly providing pertinent training examples.  While facilitating control, their effectiveness can be hindered by the lack of expertise or misconceptions. We investigate how users may conceptualize, experience, and reflect on their engagement in machine teaching by deploying a mobile teachable testbed in Amazon Mechanical Turk. Using a performance-based payment scheme, Mechanical Turkers ($N=100$) are called to train, test, and re-train a \textit{robust} recognition model in real-time with a few snapshots taken in their environment. We find that participants incorporate diversity in their examples drawing from parallels to how humans recognize objects independent of size, viewpoint, location, and illumination. Many of their misconceptions relate to consistency and model capabilities for reasoning. With limited variation and edge cases in testing, the majority of them do not change strategies on a second training attempt.
\end{abstract}





\printccsdesc

\section{Introduction}
As machine learning and artificial intelligence become more present in everyday applications, so do efforts to better capture, understand, and imagine this coexistence. Experts from diverse disciplines are working together and critically examining the impact of algorithmic decisions, their assumptions, and their biases~\cite{angwin2016machine, barocas2016big, boyd2012critical, campolo2017ai, kocielnik2019will}. Error-prone, computationally complex, and failing in ways unexpected by humans, such algorithms called early on for transparency, interpretability, accountability, and control~\cite{suchman1987plans, SWARTOUT1983285, shneiderman1997direct,  diakopoulos2014algorithmic, 10.1145/3290605.3300831}. More recently, these efforts have redoubled (surveyed in~\cite{abdul2018trends, weld2018intelligible}), fueled by funding and legal initiatives such as the DARPA Explainable Artificial Intelligence~\cite{gunning2017explainable} and the European Union's General Data Protection Regulation~\cite{eu2016gpdr}, while feeding into future initiatives such as the Algorithmic Accountability Act~\cite{AlgAccAct}.

\begin{figure}[t]
    \centering
    \includegraphics[width=\columnwidth]{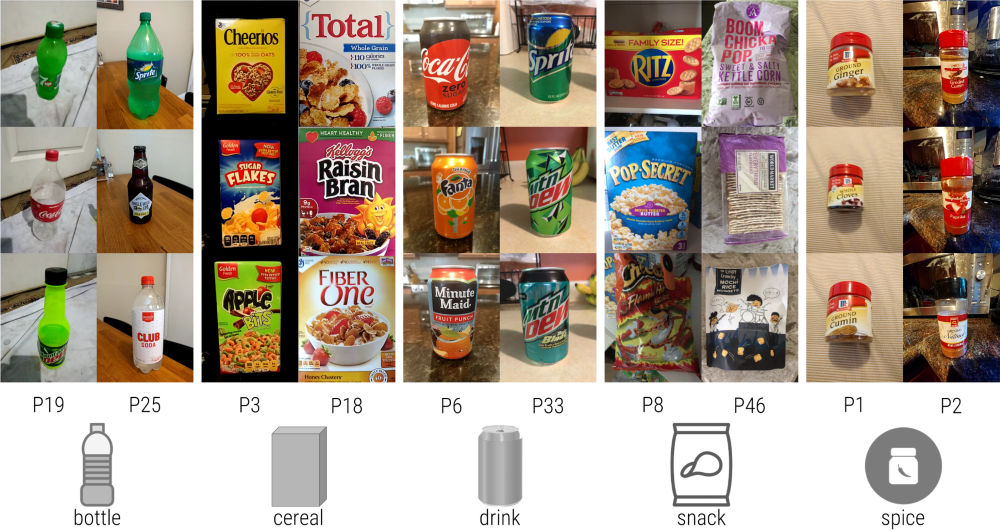}
    \caption{Given an object category, MTurkers are called to choose three object instances and train a \textit{robust} personal object recognizer using their mobile camera. Here we include examples from some of the participants' selected objects.}
    \label{fig:category_samples}
    \vspace{-0.2in}
\end{figure}

Machine teaching~\cite{DBLP:journals/corr/SimardACPGMRSVW17, zhu2018overview} lies at the core of these efforts as it enables end-users and domain experts with no machine learning expertise to innovate and build AI-infused\footnote{A term in Amershi \etal, 2019~\cite{10.1145/3290605.3300233} for ``systems that have features harnessing AI capabilities that are directly exposed to the end user.''} systems. Beyond helping to democratize machine learning, it offers an opportunity for a deeper understanding of how people perceive and interact with such systems to inform the design of future interfaces and algorithms~\cite{Amershi_Cakmak_Knox_Kulesza_2014} -- a perspective this paper shares.

Within this paradigm, teachable interfaces~\cite{patel1998teachable, GoogleTeachableMachine} explore applications where users can explicitly train a model with their generated data and labels. While facilitating user control, the effectiveness of these applications can be hindered by the lack of expertise or misconceptions about machine learning.  Though personalization is often the ultimate goal (\eg ~\cite{kacorri2017people}), the interactive nature of these interfaces can help users in return to uncover basic machine learning concepts (\eg ~\cite{hitron2019can}).

In this paper, we examine how people conceptualize, experience and reflect on their engagement with machine teaching in the context of a supervised image classification task, a task where humans are extremely good compared to machines, especially when they possess prior knowledge of the image classes. Using a teachable interface for object recognition, we recruit participants ($N=100$) through Amazon Mechanical Turk\footnote{https://www.mturk.com/} to choose three objects in their environment and train a model to distinguish between them in real-time using the camera on their mobile phones, as shown in Figure~\ref{fig:category_samples}.  

\textbf{Why crowdsourcing.} Beyond being utilized as a platform for obtaining labeled data quickly at low cost, crowdsourcing is also employed for behavioral and perception studies (\eg ~\cite{10.1145/1753326.1753357, d548a2d3f3a54c6c9400ceeb66d003e7, 10.1371/journal.pone.0051876}) including those for understanding people's interactions with machine learning systems, surveyed in~\cite{JMLR:v18:17-234}. Allowing us to quickly recruit a large participant pool for this study, it also enables data collection outside a laboratory to obtain high variability and real-world illumination, backgrounds, and camera manipulations in the user's environment. 

We build a web-based testbed for a mobile teachable object recognizer and ask participants to train and evaluate it on three objects of choice within an object category (Figure~\ref{fig:category_samples}). Categories represent daily objects that span different characteristics such as size, shape, color, material, and function. Through a performance-based payment scheme~\cite{ho2015incentivizing}, participants are called to iterate and reflect over their efforts with the goal of making their recognition models more \textit{robust}. Serving as an oracle, they are tasked with delivering a teaching set to the recognition model to help it learn the classification task. 

We conduct a contextualized quantitative analysis on the participants photos, their written responses, as well as their model performance. We find that diversity, important in machine learning, is deemed important by a majority of participants and incorporated in teaching strategies, drawing from parallels to how humans generalize across object size, viewpoint, location, and illumination~\cite{palmeri2004visual}. Many misconceptions relate to consistency; few think that it is good to be consistent and teach with almost identical examples; others failed to be consistent on incorporating diversity across classes. While participants have good intuition on importance of discriminatory features in teaching but on evaluating their models, we observe susceptibility to missing edge cases. Last, we see that the majority of participants do not change strategies on a second attempt even though possess a reasonable intuition on what would be important. We see how our findings and insights can help better understand non-experts' interactions with machine teaching and guide the design of future teachable interfaces that can anticipate users misconceptions and assumptions.

\section{Related Work}
We discuss prior work on machine teaching with a focus on teachable interfaces that most relate to our study. Prior work on behavioral studies using crowdsourcing is briefly mentioned to highlight elements that we draw from.

\subsection{Machine Teaching}
Machine teaching involves a teacher who knows the decision boundaries and designs an optimal training set for one or more students~\cite{zhu2018overview}. In this paper, the teacher is a human and the student is a classification model who is being trained to classify images of objects, as shown in Figure~\ref{fig:machine_teaching_problem_space}, though the inverse -- machines teaching humans to classify images -- is also an active area of research~\cite{johns2015becoming}. There is a rich literature on sequential machine teaching with humans as the teacher, \eg programming by demonstration for teaching robots to manipulate objects~\cite{THOMAZ2008716, DILLMANN2004109}. However, in this review we focus on prior work that utilizes batch teaching, where examples are given as a set and their order does not matter. 

Batch teaching is a very common paradigm for many real-world AI-infused systems, \eg using face recognition, fraud detection and speech recognition. This is typically done by experts in the field and end-users are hardly exposed to the underlying mechanisms that could help explain their limitations. Teachable interfaces\footnote{A term coined by Patel and Roy (1998)~\cite{patel1998teachable}, where ``the user is a willing participant in the adaptation process and actively provides feedback to the machine to guide its learning.''} that fall under this machine teaching paradigm, have the potential to help in this direction as they can enable non-experts to uncover basic machine learning concepts (\eg ~\cite{hitron2019can}). Moreover, with advances in transfer learning~\cite{pan2010survey, sun2019meta}, they can spur innovation as end-users can re-purpose models trained on vast amounts of data for new but related tasks, \eg personalize assistive technologies~\cite{10.1145/3167902.3167904}.

\begin{table}
\small
\centering
\caption{Related studies' characteristics juxtaposed with ours.}
\resizebox{8.5cm}{!}{
       \begin{tabular}{l r c c c c c c c c}
        \toprule
          & & \cite{fiebrink2011human} & \cite{6204822} & \cite{bragg2016personalizable} & \cite{kacorri2017people} & \cite{hitron2019can} &
          \cite{zimmermann2019youth} & This study \\
        \midrule
        \multirow{3}{*}{\rotatebox[origin=c]{90}{Setting}}
        & People & 1,7,21  & 10   & 12    & 8     & 30    & 5     & 100 \\
        & Controlled   & $\bullet$     & $\bullet$  & $\bullet$     & $\bullet$     & $\bullet$     & $\bullet$     &  \\
        & Real-world    & $\bullet$    &              &               &               &               &               & $\bullet$     \\
        \midrule
        \multirow{3}{*}{\rotatebox[origin=c]{90}{People}}
        & Crowd &       &       &       &       &       &       & $\bullet$ \\
        & Children &       &       &       &       & $\bullet$     & $\bullet$     &  \\
        & Disability &       &       & $\bullet$     & $\bullet$     &       &       &  \\
        \midrule
        \multirow{4}{*}{\rotatebox[origin=c]{90}{Input}}
        & Sensing & $\bullet$     &       &       &       & $\bullet$     & $\bullet$     &  \\
        & Audio &       &       & $\bullet$     &       &       &       &  \\
        & Image &       &       &       & $\bullet$     &       &       & $\bullet$ \\
        & Video & $\bullet$     & $\bullet$    &       &       &       &       &  \\
        \midrule
        \multirow{3}{*}{\rotatebox[origin=c]{90}{Output}}
        & Recognition & $\bullet$     &       &       & $\bullet$     & $\bullet$     &       & $\bullet$ \\
        & Detection &       &       & $\bullet$     &       &       & $\bullet$     &  \\
        & Control &       & $\bullet$     &       &       &       &       &  \\
        \midrule
        \multirow{3}{*}{\rotatebox[origin=c]{90}{Analysis}}
        & Accuracy &       & $\bullet$     &  $\bullet$     & $\bullet$     &       &       & $\bullet$ \\
        & Behavior & $\bullet$     &       & $\bullet$     & $\bullet$     & $\bullet$     & $\bullet$     & $\bullet$ \\
        & Feedback & $\bullet$     &       & $\bullet$     & $\bullet$     & $\bullet$     & $\bullet$     & $\bullet$ \\
    \bottomrule
    \end{tabular}%

}
\label{tab:related_work}
\end{table}

We look into prior work employing teachable interfaces, a term perhaps not originally used by the authors. Here, we focus on a subset of interactive machine learning literature, where users are called to generate all the training and testing examples for a personalized model. Table~\ref{tab:related_work} presents representative examples of prior studies from 2011-2019 on gesture recognition for musicians~\cite{fiebrink2011human}, sign language~\cite{6204822} and educational applications~\cite{hitron2019can},  personalized sound detectors for people who are deaf/Deaf or hard-of-hearing~\cite{bragg2016personalizable}, personal object recognizers for blind people~\cite{kacorri2017people}, and physical activity classifiers for young athletes~\cite{zimmermann2019youth}. In contrast to this work, prior studies tend to have smaller participant pools and are typically conducted in a controlled setting, where the researchers are present. Partially this could be due to the user characteristics of interest; people with disabilities~\cite{bragg2016personalizable, kacorri2017people}, children~\cite{hitron2019can}, and students~\cite{zimmermann2019youth}. Another reason could be challenges in remote data collection as it would require a working prototype~\cite{bragg2016personalizable, kacorri2017people} or specialized devices from the users~\cite{hitron2019can, zimmermann2019youth}. Our teachable object recognition testbed, utilizing built in camera in a mobile phone, and existing crowdsourcing platforms allow us to reach a larger participant pool that can be further scaled.

As shown in Table~\ref{tab:related_work}, the input modality for the teaching set was more often based on sensing~\cite{fiebrink2011human, hitron2019can, zimmermann2019youth} and videos~\cite{fiebrink2011human, 6204822} with one example for sound~\cite{bragg2016personalizable} and photos~\cite{kacorri2017people}. For the last two, participants could not assess the quality of their teaching examples -- participants who were deaf/Deaf or hard-of-hearing could not hear the sounds they recorded~\cite{bragg2016personalizable} and blind participants could not see the photos they took~\cite{kacorri2017people}. In this paper, we choose images as the input modality for the teaching set. This allows us to tap into a large user group of non-experts that can simply use their mobile phones to take the photos in a real-world setting. More so, by choosing an object classification task, an accessible task to many where they can serve as the oracle, we are given the opportunity to explore how humans teach a high-dimensional decision boundary to machines by feeding them only with few instances. More importantly, this modality allows us to visually inspect the teaching set for common patterns in users' behavior.

Similar to most of the prior work in Table~\ref{tab:related_work}, our analysis is based on observed behaviors and participant feedback. Leveraging prior work in neuroscience, we examine how non-experts' teaching strategies draw parallels in machine robustness to human robustness, where object recognition involves generalization across size, location, viewpoint and illumination~\cite{palmeri2004visual}. While prior work did not include such a fine-grained analysis of the participants' input, it provided insights and anecdotal evidence that guided the design of our study such as the need for iterations~\cite{fiebrink2011human, zimmermann2019youth, hitron2019can}, which may vary not only across participants but also due to the underlying algorithm and task~\cite{10.1145/604045.604056}. For comparison purposes and time sake, we opted to keep the number of iterations constant at two.  Similar to our study, the number of classes were limited (2-5) with an exception of 15~\cite{kacorri2017people}, where there were no iterations.

\subsection{Crowdsourcing and Online Behavioral Studies}
Despite the potential risks in data validity~\cite{paolacci2010running, Mason2012}, advantages such as subject anonymity, prescreening, diversity, efficiency, and low cost have made crowdsourcing platforms attractive for user studies both in social~\cite{Litman2017, PEER2017153} and cognitive~\cite{pub:101603} science with a focus on behavioral and perception studies (\eg ~\cite{10.1145/1753326.1753357, d548a2d3f3a54c6c9400ceeb66d003e7, 10.1371/journal.pone.0051876}). A building block for the machine learning community, crowdsourcing has been utilized to generate data and annotations, validate existing systems, incorporate feedback from humans, and observe how people interact with machine learning models, surveyed in ~\cite{JMLR:v18:17-234}. We build on this prior work
adopting a performance-based payment scheme~\cite{ho2015incentivizing} to incentivize participants while ensuring a rate of \$15/hour~\cite{Hara:2018:DAW:3173574.3174023}. Perhaps the closest work to our study is that of Yang \etal~\cite{yang2018grounding}, where online interviews with non-experts (N=98) were used to elicit how people with no machine learning expertise perceive machine learning processes. While the survey did not include hands-on interactions with a teachable interfaces, the findings stressed the need for future work on helping people build better learning algorithms, further motivating our work.

\section{Testbed: Teachable Object Recognizer} 

To explore how non-experts conceptualize, experience and reflect on their engagement with machine teaching, we build a web-based teachable object recognizer for mobile phones. Participants can train, test, and re-train it to distinguish between three objects of their choice. In this case, a test corresponds to a `direct' evaluation~\cite{fiebrink2011human}, where participants take photos of their objects in real-time and observe the model's behavior. To help us better contextualize our observations, participants also provide background information and feedback\footnote{Questions and prompts can be found in the supplementary material.}.

\begin{figure}
  \centering
  \includegraphics[width=\columnwidth]{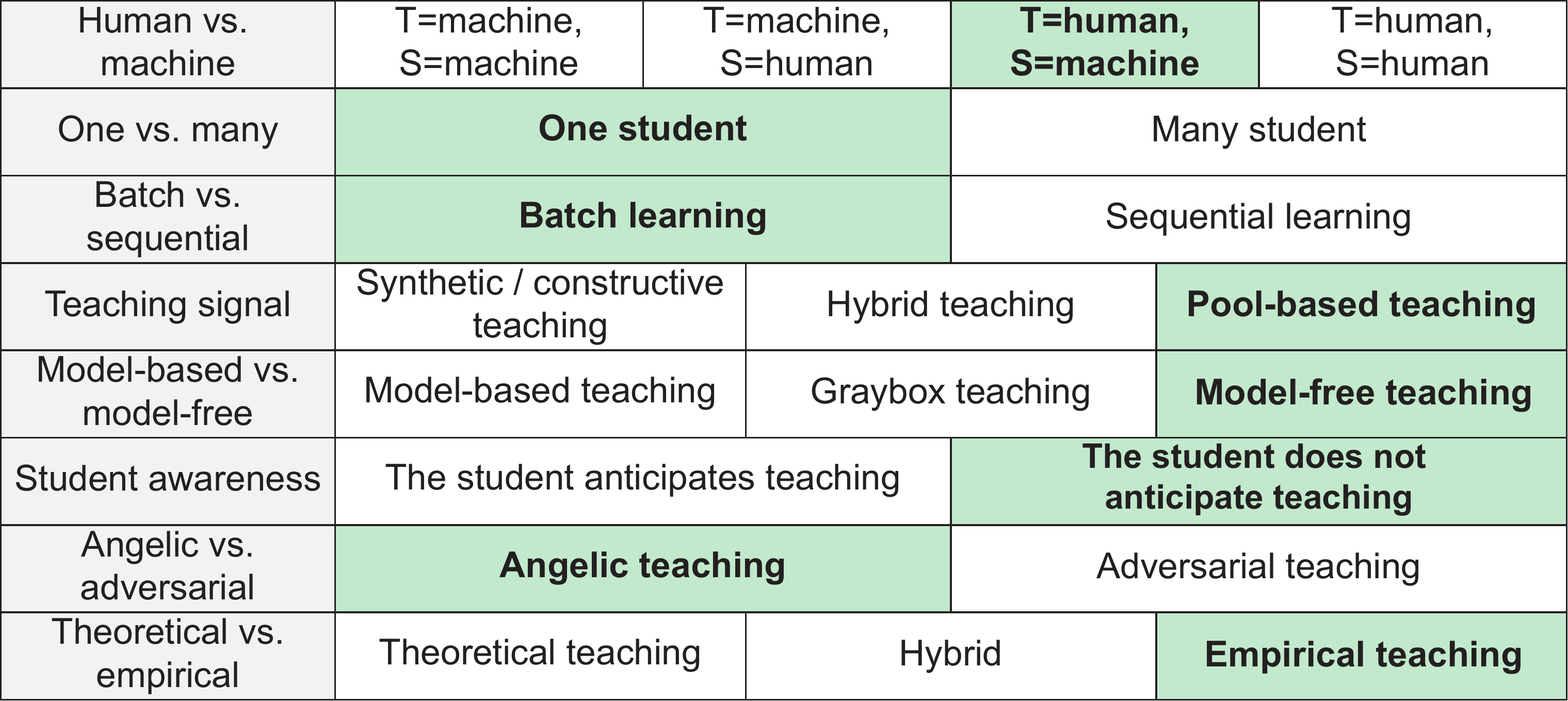}
  \caption{Characterization of our testbed in the machine teaching problem space~\protect\cite{zhu2018overview}, where T stands for teacher and S for student. A human T employs a pool-based, model-free, angelic, empirical teaching. The testbed has a single recognition model S learning in batch mode, unaware that is being taught, while considering T as a friend (no adversarial examples).}
  \label{fig:machine_teaching_problem_space}
\end{figure}

\textbf{Our machine teaching problem.} As shown in Figure~\ref{fig:machine_teaching_problem_space}, we adopt Zhu \etal~\cite{zhu2018overview} machine teaching problem space to characterize the teachable interface in our testbed as a system where human is the teacher and machine is the student. The teacher provides, in batch mode, a finite pool of examples consisting of labeled photos of objects as the teaching signal. The teacher takes a model free approach, treating the student as a blackbox, though we anticipate that humans may already have some assumptions on how the black box works or should work. The student, employing a convolutional neural network, does not anticipate teaching, \ie assuming training examples are independent and identically distributed and that there are no errors. More so, the teacher is considered a friend, \ie no adversarial training. Last, we assume that the teacher uses heuristic teaching methods to improve the performance of the student, the object recognition model in our case. We aim to better understand these heuristic methods, factors they may relate to, as well as assumptions that people may have.

\textbf{Model.} For each user, our testbed creates a new convolutional neural network using the Google Inception V3~\cite{Szegedy2016rethinking} pre-trained on ImageNet~\cite{deng2009imagenet}. Everytime the user provides a teaching set, the last layer of the pre-trained model gets replaced with a new softmax layer and re-trained with the user's images with $500$ steps and a gradient descent learning rate of $10^{-2}$. Models are trained on our 8 GPU server in real-time asynchronously; the app continues to run and ask users for open-ended feedback while the training continues in the back. The web interface communicates with the server using the Flask API~\cite{FlaskAPI}.

\textbf{Interface.} As shown in Figure~\ref{fig:screenshot}, initially the testbed asks for background information, technology experience, and familiarity with machine learning. Then, it provides five object category options: bottle, cereal, drink, snack, and spice, with three sample icons for each category indicative of the preferred shape. Categories are inspired by prior work on personal object recognizers~\cite{kacorri2017people} and are engineered to elicit objects that are present in daily life but differ in size, shape, color, material, and function. Participants can choose to train only on one of the categories. To avoid object shape or size from being a factor in any observed inconsistencies between the classes, they are asked to use objects (a total of three) that fall within the same category; three, the smallest number for multiclass classification and previously used in teachable interfaces for non-experts~\cite{GoogleTeachableMachine}, minimizes challenges in finding different object instances within a category in a real-world environment as well as the task completion time (already 40 mins long).
After labeling their objects, participants are guided through five interactions with the machine learning model (the student)\footnote{All instructions can be found in the supplementary material.}:

\begin{figure}
    \centering
    \includegraphics[width=\columnwidth]{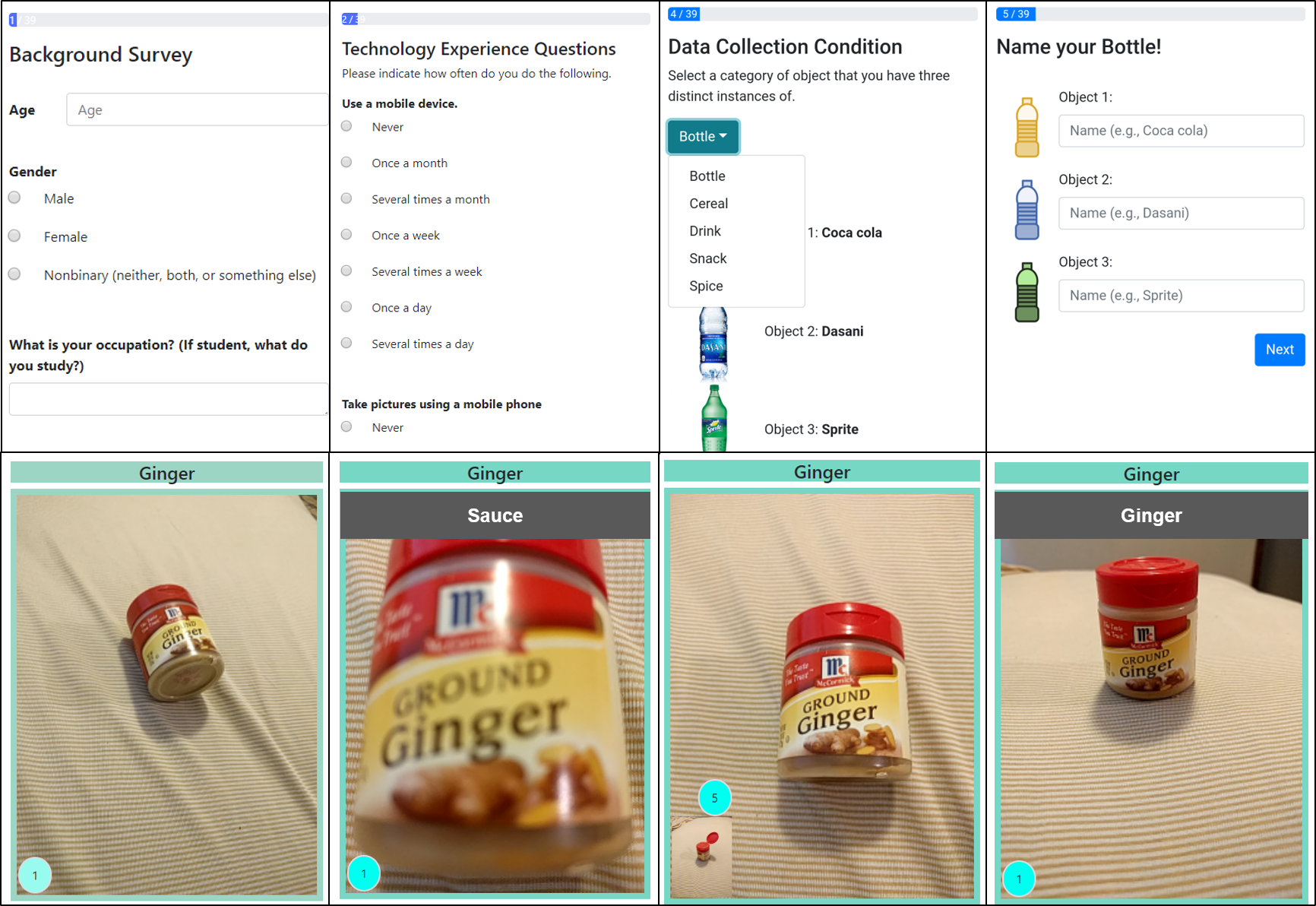}
    \caption{Testbed screenshots: questionnaires, category selection, object labeling, and camera view in training and testing.}
    \label{fig:screenshot}
\end{figure}

\textit{Preliminary test (TS0)}: Participants are asked to take photos of their objects to see if the existing non-personalized model can recognize them. The instruction reads: \textit{``Take a photo of an object (name at the top) by tapping on the camera screen. The existing model will try to predict it.''} Given an object label displayed at the top, one takes a photo of the corresponding object and sees the recognition result (a label displayed for 3 seconds). This repeats 15 times (5 times per object in a random order). As expected, during this interaction recognition results will not match participant's labels as the generic model is based on Google's Inception V3 and is not yet personalized. There is a dual motivation behind this interaction. First, it helps familiarize with the interface, which simulates the native camera app. Second, it helps collect evaluation examples unbiased from one's teaching experience that is to follow. 

\textit{Train 1 (TR1)}: Participants are asked to train the object recognizer with the following instructions: \textit{``Train our object recognizer to identify robustly your objects anywhere, anytime, for anyone. We will randomly choose one of your objects and ask you to take 30 photos of it. You will be paid \$2 extra if your examples pass our robustness test.''} Here, we hint that model robustness means to be able to recognize an object anywhere, anytime, for anyone. Motivated by Ho \etal~\cite{ho2015incentivizing} performance-based payment scheme, we also create the impression of a `secret' test distinguishing examples best for robustness, though on our end this is merely a naive quality examination (\eg photos of objects in a screen rather than in the real-world). As shown in Figure~\ref{fig:screenshot}, given an object label displayed at the top, participants take 30 sequential photos. This repeats 3 times (1 time per object in a random order). Thus, the first teaching set comprises 90 photos (30 per object).

\textit{Test 1 (TS1)}: Similar to TS0, participants are asked to \textit{``Test the trained object recognizer again to see how robust it is.''} Here, recognition labels match participants' labels except in cases of misclassification, where an object is misrecognized as one of the other two. Again, no confidence scores are shown. 

\textit{Train 2 (TR2)}: Participants are given an opportunity to re-train their model from scratch with the following instructions: \textit{``You told us what you would do differently, now show us! On the next screen, take 30 more pictures of the requested object. You will be paid \$3 extra if this training does better than the previous one in our robustness test.''}

\textit{Test 2 (TS2)}: As in TS1, users can test the re-trained model. The instruction given to the participant was \textit{``The object recognizer is trained again. Test the trained object recognizer.''}

\textbf{Eliciting Feedback.} The testbed includes the following open-ended questions: \textit{``What did you think was important to consider when training the object recognizer?''} after TR1; \textit{``If you were to retrain the system to make it more robust, what would you do differently?''} after TS1; \textit{``How did you position the object in the image?''}, \textit{``How did you decide the distance of the camera from the object?''}, and \textit{``How did you decide which side of the object is visible in the image?''} at the end.

\section{Crowdsourcing Study}
We deploy our testbed in Amazon Mechanical Turk (IRB \#1255427-1) and investigate how non-experts crowdworkers teach a machine a high-dimensional decision boundary such as a fine-grained image classification with a few examples only. 

\subsection{Participants}
\begin{figure}[b]
    \centering
    \includegraphics[width=\columnwidth]{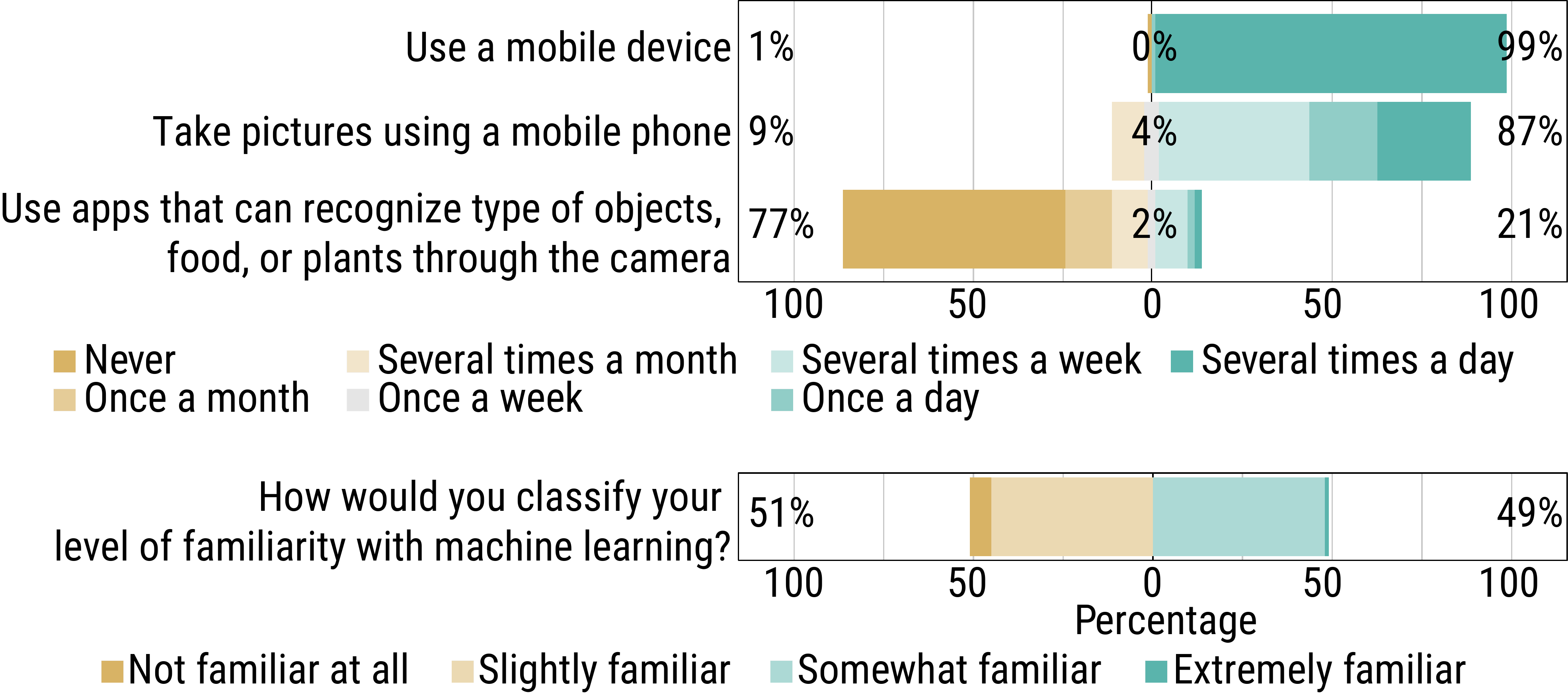}
    \caption{Participants' technology experience and familiarity with machine learning mostly ranging from slightly (have heard of it but don't know what it does) to somewhat familiar (I have a broad understanding of what it is and what it does).}
    \label{fig:chart_background}
\end{figure}
We recruited 143 participants over 10 days. However, data from 43 were excluded from the analysis -- 7 helped in piloting, 1 used the same object for all classes, 3 took photos of objects in display screens, 2 took photos with no objects. The other 30 had technical problems by attempting the task simultaneously with our system failing to distribute them across the 8 GPUs, losing data from 12 and interrupting the task for other 18; all were compensated and the bug was fixed.
The 100 participants who were included in the dataset ranged from 20 to 60 in age ($\mu$=32.6, $\sigma$=8.3);  49 were male, 50 female, and 1 non-binary with 90 reporting being right handed. No one reported a visual or motor impairment. As shown in Figure~\ref{fig:chart_background}, the majority of participants are frequent users of mobile devices taking photos with them weekly, though many of them don't use any applications for recognizing objects, food, or plants. When asked about familiarity with machine learning, 6 reported never having heard of it, 45 had heard of it but didn't know what it does, 48 had a broad understanding of what it is and what it does, and only one reported having extensive knowledge.

\subsection{Procedure} 
With the goal of attracting non-experts in machine learning, we opted for a HIT description that minimizes technical terms:
\noindent \textit{``You will be asked to take photos of everyday products such as soda cans, cereal boxes, and spices to teach your phone to automatically recognize them. To see how well the object recognition works you will test it by giving a single photo at the time.''} A warning message was displayed if participants attempted to start the study from a device other than a mobile phone. Only one participation was allowed. 

Through piloting, we estimated that a study session could be successfully completed within 30-40 minutes. Adopting a \$15/hour compensation rate~\cite{Hara:2018:DAW:3173574.3174023} all participants received a total of \$10 once all the data collection was completed. To incentivize participants, we used a performance-based payment scheme~\cite{ho2015incentivizing}, where this amount was split as \$5 flat participation, \$2 bonus for passing \textit{``our robustness test''} in the first attempt to train, and \$3 bonus for achieving a better performance in \textit{``our robustness test''} the second time around. Given that objects differ across participants it was not possible to have an ideal \textit{`secret robustness test'}; bonus was decided merely on a quality check.  While the testbed's connection is persistent and one could do other tasks in between, we observe that participants took on average 35.57 minutes (14.21-79.86, $\sigma$=12.85) to complete the study, very close to our estimates.

\begingroup
\setlength{\tabcolsep}{6pt} 
\renewcommand{\arraystretch}{1.1} 

\section{Analysis of Behavior}
We explore how participants conceptualize, experience and reflect on their engagement with machine teaching by looking at the photos they took for the teaching and testing sets as well as changes in their behavior when repeating the process. 
Observations are contextualized with participants' responses.

\textit{Visual Attributes in Photos.} We collected a total of $22,500$ photos from 100 participants across all training and testing interactions. To uncover patterns in participants' teaching strategies, photos were coded using thematic coding~\cite{braun2006using}. Two researchers independently created initial codebooks of visual attributes in photos across four dimensions, \ie size, location, viewpoint, and illumination; prior work on visual object understanding~\cite{palmeri2004visual} indicates that our ability to recognize objects generalizes across these dimensions. We want to see how participants draw parallels from their understanding of robustness in these dimensions to enable machines to do the same.

Researchers discussed disagreements to produce a final codebook, shown in Tables~\ref{tab:codebook_binary}--\ref{tab:codebook_presence} with examples in Figures~\ref{fig:samples_variation} and~\ref{fig:samples_presence}. There are two types of attributes: binary and count. Binary attributes capture presence of variation or inconsistency within a teaching or testing set of photos. If a participant varied photos for an object along an attribute such as distance (\textit{VSizeDist}) or background (\textit{VLocBg}), the corresponding attribute is 1; otherwise 0. Similarly, variation inconsistency across the three objects is captured through binary attributes, named \textit{ISize, ILoc, IView, IIllum}. Count attributes indicate the number of photos within a set with a certain characteristic such as presence of participant's hand (\textit{CHands}) and use of flashlight (\textit{CFlash}) or a quality issue such as dark (\textit{QDim}) and blurry (\textit{QBlurry}) photos.  There was substantial agreement
 (Cohen's kappa=0.80).
 
 \begin{figure}
    \centering
    \includegraphics[width=\columnwidth]{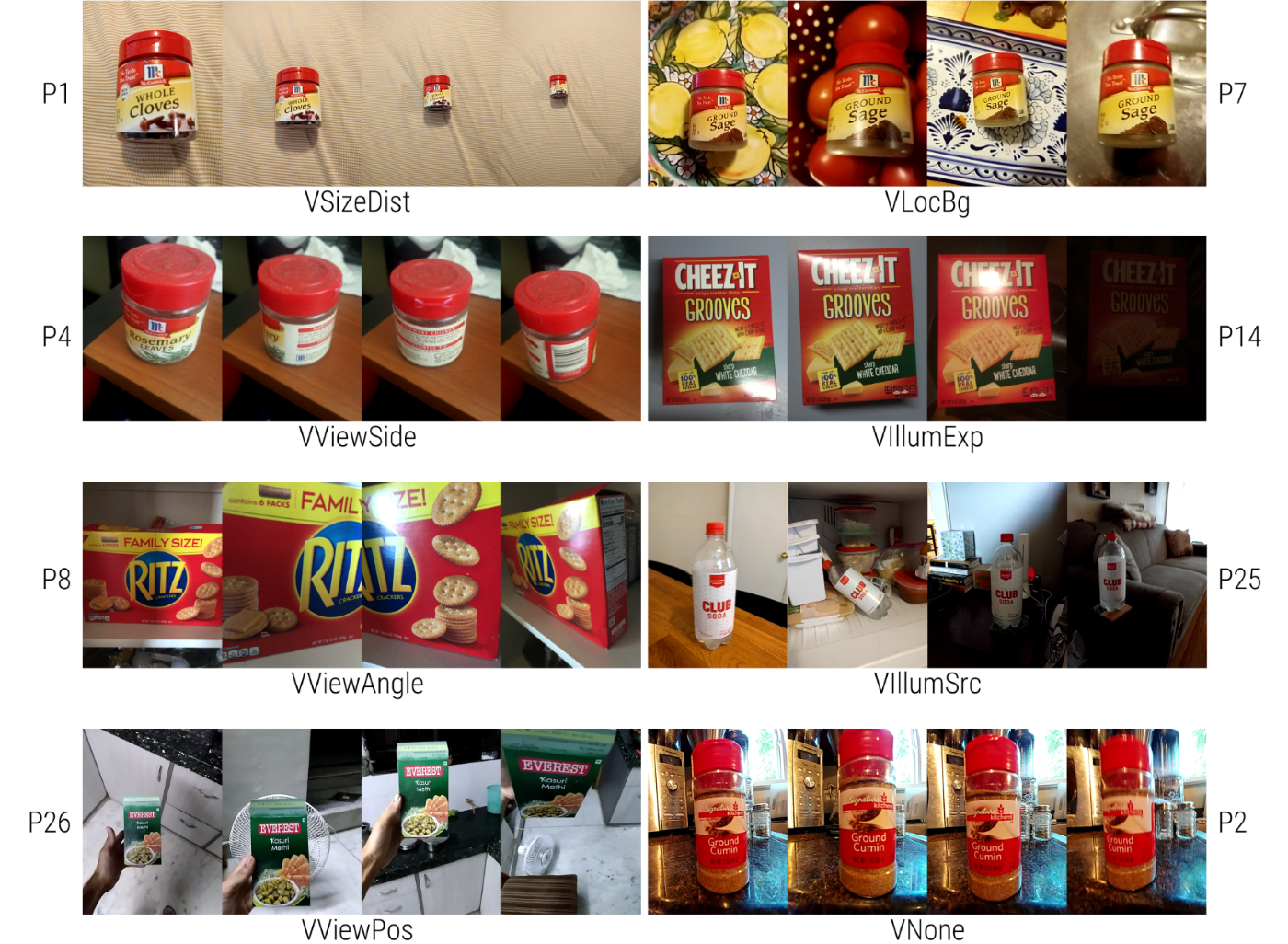}
    \caption{Examples of variation attributes in teaching sets.}
    \label{fig:samples_variation}
    \vspace{-0.1 in}
\end{figure}

\begin{figure}
    \centering
    \includegraphics[width=0.85\columnwidth]{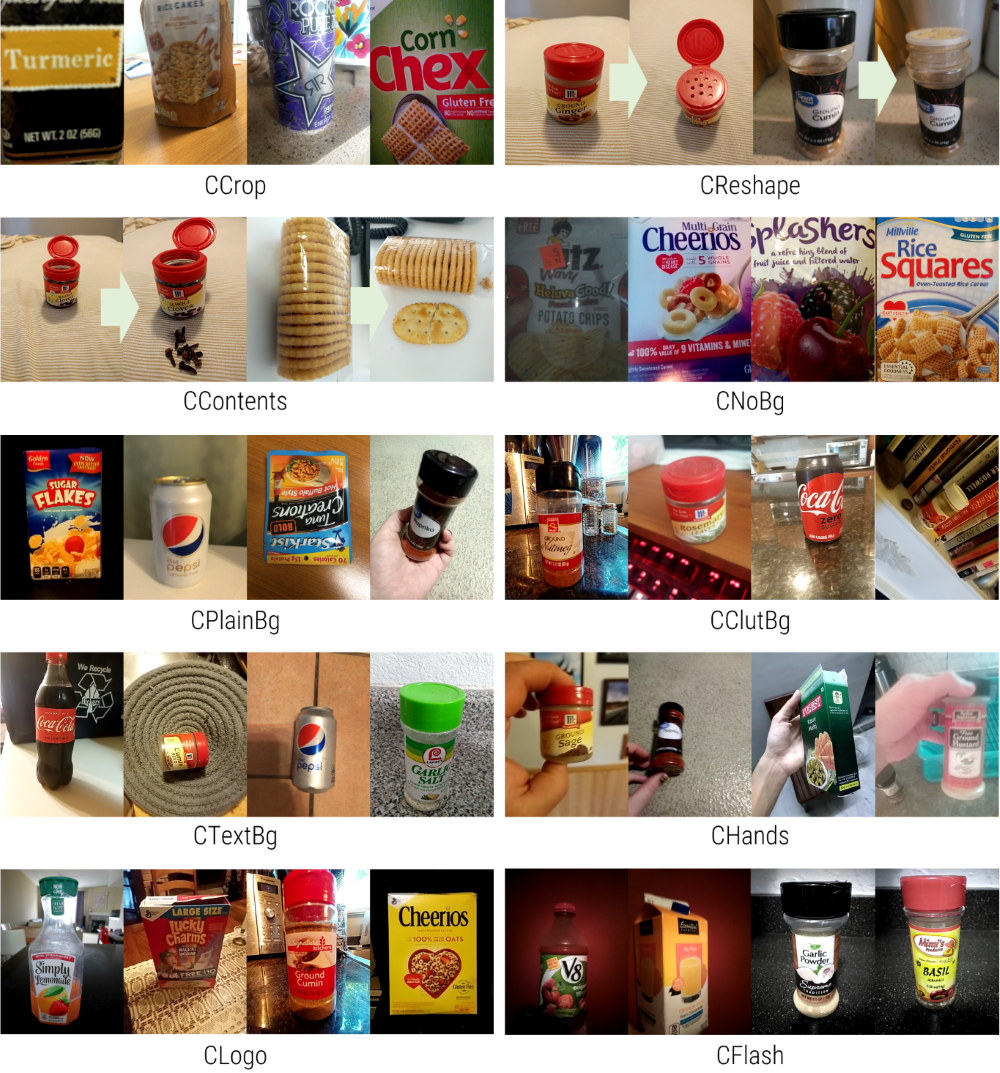}
    \caption{Sample photos considered by the count attributes.}
    \label{fig:samples_presence}
\end{figure}

\begin{table*}
   \small
  \centering
  \caption{Variation attributes, true if a variation is present for at least one object.}
    \begin{tabular}{m{1.2cm}|m{15.8cm}}
    \textbf{Variation}& \textbf{Definition} \\ \Xhline{2\arrayrulewidth}
    \textit{VSizeDist} & True if camera \textbf{distance}, ratio of object height to frame, differs for two or more photos using [0, 0.25), [0.25, 0.5), [0.5, 1.0), and  [1.0, $\infty$) bins.\\ \hline
    \textit{VLocBg} & True if the \textbf{background} differs for two or more photos, \ie different locations or perspectives of a space. \\ \hline
    \textit{VViewSide} & True if the \textbf{side} of objects differs for two or more photos. \\
    \textit{VViewAngle} & True if the \textbf{angle} between the camera and the object with the same side of object differs for two or more photos. \\
    \textit{VViewPos} & True if the \textbf{position} of the object in the camera frame, center, top left, top right, bottom left, or bottom right, differs for two or more photos. \\ \hline
    \textit{VIllumExp} & True if the \textbf{exposure} to light differs for two or more photos taken at the same location. \\
    \textit{VIllumSrc} & True if the \textbf{source} of light differs for two or more photos because they were taken at different locations. \\ 
    \end{tabular}%
  \label{tab:codebook_binary}%
\end{table*}
\endgroup

\begingroup
\setlength{\tabcolsep}{6pt} 
\begin{table*}
    \small
  \centering
  \caption{Inconsistency attributes, true if there is an inconsistency in variation across the three objects.}
    \begin{tabular}{m{1.5cm}|m{15.5cm}}
    \textbf{Count} & \textbf{Definition} \\ \Xhline{2\arrayrulewidth}
    \textit{ISize} & True if the camera \textbf{distance} varies in the photos for one or two objects but not all three. \\ \hline
    \textit{ILoc} & True if the \textbf{background} varies in the photos for one or two objects but not all three. \\ \hline
    \textit{IView} & True if size, angle, or position capturing \textbf{viewpoint} varies in the training photos for one or two objects but not all three. \\ \hline
    \textit{IIllum} & True if light exposure or source capturing \textbf{illumination} varies in the training photos for one or two objects but not all three.
    \end{tabular}%
  \label{tab:codebook_inconsistency}%
\end{table*}%
\endgroup

\begingroup
\setlength{\tabcolsep}{6pt} 
\begin{table*}
    \small
  \centering
  \caption{Count attributes, number of photos with a given characteristic including those looking at quality issues.}
    \begin{tabular}{m{1.5cm}|m{15.5cm}}
    \textbf{Count} & \textbf{Definition} \\ \Xhline{2\arrayrulewidth}
    \textit{CCrop} & Number of photos where the object is \textbf{cropped}, \ie object is close to the camera, out of frame, or obscured by another object. \\ \hline
    \textit{CReshape} & Number of photos where the object was \textbf{reshaped} (\eg opening a lid of a package). \\ \hline
    \textit{CContents} & Number of photos where the \textbf{contents} inside a package was taken out of the container or the inside of the package is visible. \\ \hline
    \textit{CNoBg} & Number of photos where were the background is \textbf{not visible} because the photos are filled with the object completely. \\ \hline
    \textit{CPlainBg} & Number of photos where the background includes two or fewer colors with \textbf{no or very simple textures}. \\ \hline
    \textit{CClutBg} & Number of photos where the background is \textbf{cluttered} with objects other than the object of interest. \\ \hline
    \textit{CTextBg} & Number of photos where the background includes a wall, floor, or furniture with \textbf{texture}. \\ \hline
    \textit{CHands} & Number of photos where the participant's \textbf{hand(s)} is visible in the photo. \\ \hline
    \textit{CLogo}  & Number of photos where the side with the \textbf{logo} (or label) of the object was visible in the photos. \\ \hline
    \textit{CFlash} & Number of photos where the brightness varies in different parts of the photo like using \textbf{flashlight}. \\  \Xhline{2\arrayrulewidth}

    \textit{QSmall} & Number of photos where the object is too \textbf{small} (height of the object < 25\% of the height of the photo).\\ \hline
    \textit{QDim}        & Number of photos where the brightness of the photo is too \textbf{dark} to recognize texture or edge of the object.\\ \hline
    \textit{QBlurry}     & Number of photos where the object of interest is \textbf{blurry}.\\ \hline
    \textit{QIrrelevant}  & Number of photos where the photo includes only \textbf{irrelevant} objects without the object of interest.
    \end{tabular}%
  \label{tab:codebook_presence}%
\end{table*}%
\endgroup

\begin{figure*}
    \centering
    \includegraphics[width=2.0\columnwidth]{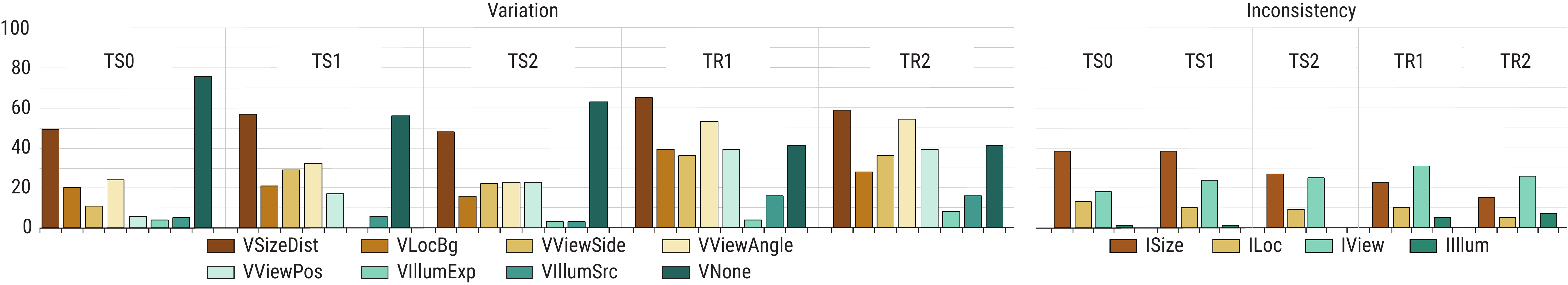}
    \caption{Number of participants per variation and inconsistency attribute across all five interactions with the model: preliminary test (TS0), train 1 (TR1), test 1(TS1), train 2 (TR2) and test 2 (TS2). The graphs on the left indicate how participants incorporate diversity in their photos in terms of object size, viewpoint, location, and illumination when they train and debug their models. 
    }
    \label{fig:chart_variation_inconsistency}
\end{figure*}

\begin{figure*}
    \centering
    \includegraphics[width=2.0\columnwidth]{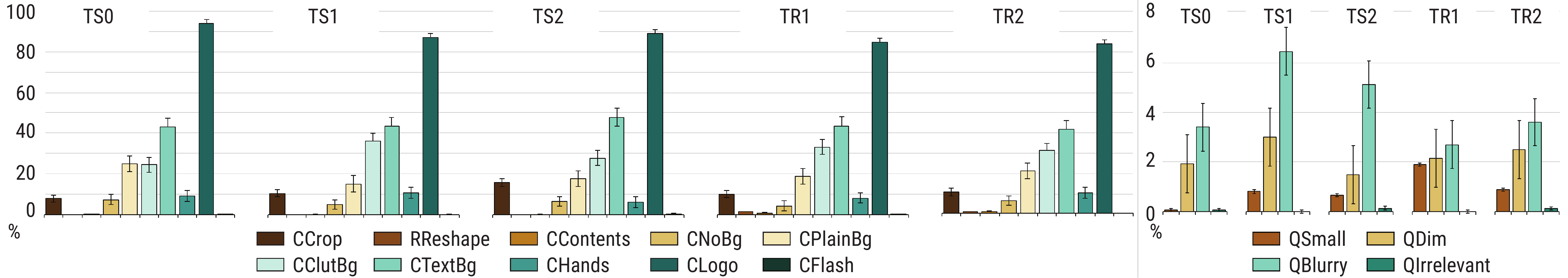}
    \caption{Percentage of photos per participant given a count attribute, with standard error as error bars. Participants took photos mostly with the logo on it and many of them against a textured or cluttered background. Often the objects were cropped in the camera frame and sometimes participants' hands were included in the photos. Surprisingly, few participants opened the object and trained the model on their content as well. The most common quality issues were blurry and dim photos though not that prevalent.
}
    \label{fig:chart_count}
\end{figure*}

\textit{Subjective Feedback.} Participants' responses to the open-ended questions were also analyzed with a thematic coding approach~\cite{braun2006using}. The same two researchers who coded the photos, created initial codebooks and merged them through discussions resolving disagreements. Responses were coded independently with a substantial agreement (Cohen's kappa=0.73).

\subsection{Results}
\subsubsection{What Are Non-experts' Teaching and Debugging Strategies?}
We explore how variation\footnote{A preliminary analysis of this appears in a work-in-progress~\cite{hong2019exploring}.}, inconsistency, and other attributes manifest on participants' image sets when they are first called to train the object recognizer on objects of their choice.

\textbf{Incorporating diversity in teaching.} Diversity plays an important role in machine learning~\cite{8717641}. When incorporated in the teaching set, it ensures that examples can provide more discriminatory information to help the model learn. By looking at participants' photos (results in Figure~\ref{fig:chart_variation_inconsistency}) and by reading their responses, we find that the majority of the participants share this intuition, but not all. In detail, 23 participants (age 21--60, $\mu$=37.57, $\sigma$=9.87) did not include any kind of variation in their TR1 teaching set --  3 of them reported never having heard of machine learning, 12 had heard of it but did not know what it does, and 8 had a broad understanding of what it is and what it does.  Immediately after training, when asked about what they considered important, 5 participants referred to the need for consistency, which in this context contradicts the way machines and people learn. For instance, P6 said \textit{``I figured I needed to be consistent when I took the picture so they looked similar.''} and P30 \textit{``Keeping the pictures the same.''} Others, who did not consider this type of consistency, mentioned that it is important to have a good quality photo where the object is well framed (4) with visible labels (8) and images that are clear (6) with ample light (2). Without even having tested their model, P2 said: \textit{``Getting different angles and perspectives so the trainer could recognize it more easily''} -- a contradiction to their initial teaching set that had no variation. We observed that in TR2, P2 reflected on this observation and varied both the object size and viewpoint. Only two other participants from this group did so as well, P5 and P18. They said having the ``name and color in'' is important in TR1 but also varied the camera distance (P5) and angle (P18) in TR2.

However, the majority of participants ($N=77$) diversified examples in their first attempt. They varied either size ($N=65$) or viewpoint ($N=63$), with some considering location ($N=39$) and illumination ($N=19$). Light exposure was least diverse ($N=4$). Looking at responses on important considerations for training, many participants ($N=52$) mentioned these strategies\footnote{All questions, instructions, and prompts prior to training were carefully edited not to prime participants towards our coding attributes.} and reflected on the need for diversity with concrete terms such as \textit{``different'', ``various'', ``all'', ``many'', ``multiple'', ``every'', ``variety'', and ``difference''} combined with \textit{``angles'', ``views'', ``sides'', ``facets'', ``background'', ``lighting'', ``distance'', and ``positioning''}. These terms correspond to the four dimensions of our coding scheme informed from prior work on visual object understanding~\cite{palmeri2004visual}, highlighting that humans' strategies for machine teaching parallel their own abilities. However, only 11 participants (age: $\mu=34, \sigma=8.71$) incorporated diversity in their teaching set across all four dimensions -- 3 reported having heard of machine learning with no further understanding, and 8 had a broad understanding of what it is and what it does.

\textbf{Being fair and consistent between classes.} Model consistency across classes is a desirable trait in machine learning with many social implications for fairness, whose definition is still being debated in the community (\eg~\cite{narayanan2018fat, mehrabi2019survey}).  There is anecdotal evidence on non-experts learning to balance class proportions in the training set over multiple iterations~\cite{fiebrink2011human, zimmermann2019youth}. By keeping the number of training examples constant, we look into their behavior across other potential disparate treatments.
Given that many participants considered diversity important for good performance, we explore how fair\footnote{In this work classes are object instances that fall within the same category and consequently share similarities such as shape, size, and material in the context of the decision making task of incorporating variation. Thus, we consider ``individual fairness''~\cite{dwork2012fairness}, where ``similar individuals should be treated similarly'', and explore whether object instances within a category are being treated the same by a participant when introducing variation in the training photos.} (\ie consistent) they are in incorporating diversity across their three objects, with results shown in Figure~\ref{fig:chart_variation_inconsistency}.  Beyond the 23 participants who did not introduce any variation for any object, we find that there were 30 other participants that were consistent. This is promising, especially since this included participants from all levels of familiarity with machine learning: not familiar at all ($N=1$), slightly familiar ($N=11$), somewhat familiar ($N=17$), and the only participant in our study that reported being extremely familiar ($N=1$). While none of these participants explicitly mentioned consistency as important, we find that more than half of them ($N=16$) continued doing so in their second attempt at training, in TR2. For the remaining 47 participants, their inconsistencies were found in variations related to all four dimensions: object size ($N=21$), viewpoint ($N=31$), location ($N=10$), and illumination ($N=5$).  

\textbf{Deciding what to show in the teaching set.}  We analyze the fine-grained count attributes in teaching and training sets (Figure~\ref{fig:chart_count}) to uncover common teaching patterns across participants.  Khan \etal~\cite{khan2011how} observed that one of the most prominent teaching strategies for a binary classification task among non-experts, called the \textit{extreme} strategy, is consistent with the ``curriculum learning'' principle~\cite{bengio2009curriculum, lee2011learning}, where participants start with the most extreme examples and continue with those closer to the decision boundary\footnote{In the Khan \etal~\cite{khan2011how} study participants did not generate the examples but they ordered them as most representative of the two classes and chose to teach one by one using all of them or a subset.}.  While our batch teaching task does not allow for a similar sequential analysis, we find that almost all participants ($N=98$) included the logo (or label) of objects in their teaching sets; on average 84.9\% ($SD=25.0$) of any participants' images included logos. This indicates that participants understand that logos and labels tend to include the most discriminatory features, which serve as the most extreme examples. Then, through variation they add less discriminative viewpoints that are closer to the decision boundary. Indeed, 18 participants explicitly mentioned logos or labels being important in training. For instance, P36 said \textit{``... trying to have a constant label view''} and P46  \textit{``... a clear shot of the front of the package with minimal background interference.''} When looking deeper at these responses though, we find that many of the participants assumed that the machine would read the text. For example, P28 said \textit{``It [the model] recognizing the different cereals by name''} and  P44 \textit{``Getting a clear shot where the writing and the size are clear.''}

In terms of the background, we find that the majority were textured ($N=66$) or cluttered ($N=62$), while many used plain ($N=48$) and a few none at all ($N=11$) -- the latter two are preferred since very few varied the object location. We observe that 26 participants included their hands in the photos. The presence of hands has been leveraged to better distinguish objects by modeling the contextual relationship between grasp types and object attributes~\cite{cai2018understanding} or to estimate the object of interest in a clutter environment~\cite{lee2019revisiting, lee2019hands}. However, given this study's fine-grained task, the grasp is expected to be similar across object of the same category. Thus, the presence of the hand doesn't really help, especially if it is not applied consistently across classes. More surprisingly, we observe that 8 participants reshaped their objects, \eg opened the lid, and 4 decided to train on the content of the object as well, \eg cinnamon powder. When asked what is important for training, one of these participants, P76, said: ``Getting lots of different angles and different ways the spice could be portrayed.'' In general, there were not many photos with quality issues. Participants took clear photos in most cases and many of them mentioned the importance of image quality in their responses, but some ($N=36$) mistakenly took a few blurry photos. Also, objects sometimes appeared too small ($N=17$) and occasionally the light was dim ($N=9$).

\textbf{Debugging and including edge cases in testing.} When asked to evaluate their model in TS1, many participants ($N=30$)
did not diversify their images at all -- 2 of them reported never having heard of machine learning, 17 had heard of it but didn't know what it does, and 11 had a broad understanding of what it is and what it does. This means that they did not check whether the recognizer is robust. We also find that compared to training, fewer participants diversify their testing set across object size ($N=57$), viewpoint ($N=49$), location ($N=21$) and illumination ($N=6$). This could be explained by many factors such as: smaller number of photos in testing (15) compared to training (90); difficulty in conceptualizing robustness; assumptions about machine's generalizing capabilities; not anticipating future uses of the model under different circumstances; or simply minimizing efforts for this HIT.  Logos were still included by the majority of the participants ($N=98$) and the same number of participants ($N=11$) took photos that did not include any background, keeping their testing data consistent with their training examples. Similar to what Zimmermann \etal~\cite{zimmermann2019youth} observed, participants ``enacted [testing] practices wherein their models appeared to have high reliability but questionable validity.'' We also find that participants took fewer photos with plain background ($W=756$, $Z=2.17$, $p=.030$, $r=0.15$), and objects that were too small ($W=126.5$, $Z=2.61$, $p=.011$, $r=0.18$) using a Wilcoxon signed rank test. None of the interesting object reshaping, or content images present in training, carried over to testing; a similar behavior to Kacorri \etal~\cite{kacorri2017people}, with ``exaggerated'' variation in training unobserved in testing.


\subsubsection{Do Teaching Strategies Evolve Through Iteration?}
Prior work indicates that the interactive nature of teachable interfaces can help users uncover machine learning concepts~\cite{hitron2019can}. We ask participants whether they would do something differently were they to retrain the model for a second time and offer a bonus if they could make it even more robust.

\textbf{Updating teaching strategies to improve performance.} ``Is this information a signal or noise'' was  one of the most common debug strategies by experts~\cite{yang2018grounding}. We investigate whether participants employ a similar approach by comparing TR2 to TR1 in terms of the variation, inconsistency, and other image characteristics, which serve as information signals for the model. Using a McNemar test for binary and Wilcoxon signed rank test for count attributes, we find the only significant difference is variation of location as observed by changes in the photo background (VLocBg). More participants diversified the background in their teaching set on the first attempt than the second ($\chi^2(1, N=100)=4.35, p=.037, \phi=0.21$, the odds ratio is 11.86). As in Zimmermann \etal~\cite{zimmermann2019youth}, we suspect that participants were trying to maximize performance by increasing consistency between their training and testing data, even though in our prompts we had defined robustness as ability to recognize the objects anywhere, anytime, for anyone. No other significant differences were observed, though this could be partially explained by limitations in the binary nature of our variation and inconsistency attributes failing to capture changes in magnitude. We shed light into other possible explanations by looking at participant's responses.      

When asked about what they would do differently if they were to retrain, some ($N=22$) said \textit{"nothing"}, \textit{"wouldn't do it differently"}, and \textit{"would not change anything"}. Few said they had nothing to change because they were satisfied with the performance in TS1 ($N=6$). For instance, P23 said \textit{"Nothing it seems very robust after the learning phase."} This was not a surprise given that in TS1 participants did not opt for a thorough evaluation, as discussed above.  \textit{"Having no idea what to change"} was also mentioned by some ($N=19$) reflected by terms such as \textit{"not sure"}, \textit{"unsure"}, \textit{"I can't think of anything"}, \textit{"have no idea"}, or \textit{"don't know"}. Indeed, 
we find that the models of these 22 participants perform well on their own test data with an average $F_{1}$ score of 0.981 ($SD=0.048$)\footnote{Only recognition labels are available in testing and no scores.} and significantly better than the rest of the participants ($U=1472, Z=5.22, p<.001, r=0.52$); a trend that carries over to the second attempt. 

Few participants wanted to change elements of the teaching process such as improving the testbed ($N=3$), taking photos faster ($N=1$), adding more classes ($N=2$), or adding more samples ($N=6$). Yang \etal~\cite{yang2018grounding} characterized the latter as ``most non-experts' only strategy to improve a model's performance.'' Others focused on improving the quality of their teaching set such as better focus ($N=5$), more light ($N=2$), show labels ($N=2$), better framing  with a certain distance ($N=1$), and centering ($N=1$). Few participants ($N=2$) explicitly mentioned the importance of the background, with P83 saying \textit{``I would try to change the color of the background to ensure that it knows what the actual object is. I think it was confused by the curry because of the black stove background which may look like the black cap of the cumin.''} Surprisingly, one participant (P85) pointed to discriminatory limitations of their objects uncovering challenges in fine-grained classification by stating \textit{``Change objects to not look so similar.''}

Last, some participants ($N=22$) explicitly indicate that adding more variation in their training set is something they would do. For instance, P14: \textit{``I would take a wider variety of angles''} and P21: \textit{``Take picture from many different locations lighting and positions.''} Only one, P36 mentioned doing so in testing, \textit{``Test different sizes''}.  When examining what they actually did in their second attempt at training, we find differing approaches: some indeed started incorporating new variations ($N=13$), some perhaps changed the magnitude as variations were present in both first and second attempt ($N=5$), and others ($N=4$) did not make those changes. While variation for these 22 participants was mostly limited to the 4 dimensions (size, viewpoint, location, and illumination), few other participants ($N=5$) indicated that they would also include different forms of the same object, \eg different containers, perhaps difficult within this study.

\section{Analysis of Performance}
We report the performance of the models that the participants train by looking at the predicted labels during the first and second round of testing using the $F_1$ score measure (F-score).

\textbf{Relating observed behavior to performance.}
 Participants achieved on average a 0.75 ($SD=0.38$) F-score in their first attempt to train the model. Using a multiple linear regression, we explore how attributes capturing their behavior in teaching and testing may relate to the relative performance of their models. While this performance is far from an ideal controlled robustness\footnote{Such a neutral test is unrealistic in our study since participants choose different objects in different environments.}, it can provide some context for the observations above such as participants' behavior in the second attempt. We use a square root transform of the F-score\footnote{Transformation is used to meet the normality assumption.} as the dependent variable. As independent variables, we use variation, inconsistency, and count attributes in TR1 and TS1 and their interaction. For model selection, we use stepwise variable selection based on Akaike information criterion (AIC)~\cite{1100705} with results shown in  Table~\ref{tab:multiple_regression}. We find that only 28\% of the variability in recognition performance is accounted by this model, as indicated by the adjusted R-squared metric. While this is modest, it is not surprising, as there are many factors that can contribute to the performance of an image classification algorithm. For instance, performance can vary based on object similarities, a common challenge in fine-grained classification; a similarity that is not directly captured by our attributes. 
 
 \begin{table}
\centering
\small
    \begin{tabular}{lllll}
    \Xhline{2\arrayrulewidth}
    \textbf{Attempt}        & \textbf{Variable}        & \textbf{Estimate} & \textbf{Std. Error} & \textbf{t value} \\ \hline
                            & (Intercept)       & 0.939            & 0.048              & 19.79***         \\
    \multirow{4}{*}{TR1} & VIllumExp            & 0.167            & 0.063              & 2.64**           \\
                            & VIllumSrc         & -0.076           & 0.049              & -1.55            \\
                            & CCrop             & 0.000            & 0.002              & 0.12            \\
                            & CPlainBg          & -0.002           & 0.001              & -2.50*           \\ 
                            & CTextBg           & -0.001           & 0.001              & -1.55            \\ \hline
    \multirow{4}{*}{TS1}    & VSizeDist         & -0.068           & 0.037              & -1.81.           \\
                            & VViewSide         & 0.108            & 0.038              & 2.83**           \\
                            & VViewPos          & -0.089           & 0.045              & -1.97.           \\ 
                            & CCrop             & 0.048            & 0.012              & 4.04***          \\ 
                            & CClutBg           & -0.007           & 0.003              & -2.14*           \\ 
                            & QBlurry           & -0.016           & 0.009              & -1.74.           \\ \hline
    TR*TS                   & CCrop             & -0.001           & 0.000              & -3.16**     \\ \Xhline{2\arrayrulewidth}
    \multicolumn{5}{c}{Signif. codes:  0 '***' 0.001 '**' 0.01 '*' 0.05 '.' 0.1 ' ' 1} \\
    \multicolumn{5}{c}{Residual standard error: 0.157 on 87 degrees of freedom} \\
    \multicolumn{5}{c}{Multiple R-squared:  0.3681,	Adjusted R-squared:  0.2809} \\
    \multicolumn{5}{c}{F-statistic: 4.223 on 12 and 87 DF,  p-value: 3.195e-05} \\
    
    \end{tabular}
    \caption{Modeling recognition performance based on attributes capturing variation, inconsistency, and other characteristics. }
    \label{tab:multiple_regression}
\end{table}
 
 In training, we find that variation in light exposure (VIllumExp) relates positively with the F-score, though very few participants included this type of diversity in their teaching set.
 We also see that the number of images where the object is taken against a plain background (CPlainBg) has a negative relationship with model performance. Though counter-intuitive, we suspect that  lack of diversity in the background might have contributed to a model that does not generalize well, \eg when tested. This seems to be supported by the negative relationship of the number of cluttered background images during testing.  
 
 In testing, we find that variation in object size (VViewSide) relates positively with the F-score.
 We also see that the number of images where objects appear to be cropped (CCrop) has a positive relationship with model performance. A plausible explanation could be that these attributes capture participants' behavior of zooming in on the object's most discriminative features, thus helping the model to distinguish objects. However, when considered as an interaction between training and testing (TR1*TS1-CCrop), this attribute appears to be negatively related to the model performance perhaps pointing to the sensitivity for consistency between the two -- if you crop objects in one case, then it helps to do so in the other as well.

\textbf{Improving performance the second time around.} As shown in the previous analysis, we observe few changes in participants' teaching strategies in the second training as captured by our attributes -- though some participants said they would do things differently. We find that this is also reflected when comparing the performance of their second model to the first. On average, participants achieved a 0.746 ($SD=0.38$) F-score the first time and a 0.749 ($SD=0.28$) the second with no significant change ($W=80.5, Z=-0.16, p=.871$).  However, participants who indicated they would do nothing to improve their model after the first attempt ($N=22$), seem to achieve significantly higher performance than the rest ($U=1472, Z=5.22, p<.001, r=0.52$) and this is a consistent trend across both attempts ($U=1459.5, Z=5.12, p<.001, r=0.51$).
Looking at these relative low F-scores for such a simple 3-way classification task, it is surprisingly that the second group of participants did not further improve their performance even though they expressed reasonable strategies. Perhaps the incentives were not strong enough and they had a higher threshold for errors, or there was not enough time and iterations to try things out. It could simply be that their object instances were too similar. Indeed, the majority ($N$=38) of the participants in this group had chosen spices.

\section{Discussion}
We see how our results, some being new insights, others strengthening prior empirical and anecdotal evidence, can help better understand non-experts’ interactions with machine teaching and guide the design of future teachable interfaces. We highlight some of them with the following suggestions:

\textbf{Account for teaching strategies}:  Our observations suggest that non-experts mainly tend to teach with clear representative examples and sometimes incorporate examples that are closer to the decision boundary through variation, which draws from parallels to how humans generalize for similar recognition tasks. In the case of object recognition, these were object size, viewpoint, location, and illumination~\cite{palmeri2004visual}; though all four were considered only by a few. Our analysis also suggest that beyond class imbalance~\cite{fiebrink2011human, zimmermann2019youth}, there can be other disparate treatments such as inconsistency in the way variation is incorporated across classes. 

\textbf{Anticipate misconceptions}: A prevalent misconception relates to consistency. While it is true that consistency between training and testing data will result in better performance, assuming they both represent real-life examples, some thought that being consistent entails teaching with multiple identical examples with no variation whatsoever. Other misconceptions relate to the capabilities of the machine for reasoning. For example, participants would train with visually disparate examples from both the container and its content separately. Others would assume that the models were able to infer the text.

\textbf{Help users craft evaluation examples}: Our observations indicate that testing examples tend to be less diverse or not at all. Thus, it is no surprise to see many people wanting to change nothing, being satisfied with the performance, or not knowing what to do. Even those who did change their behavior when training for a second time, it was to not vary the background rather than making their model more generalizable.  Help may look different based on the goal of the teachable interface. If it is personalization (\eg \cite{kacorri2017teachable}), then it could mean guiding the user to generate examples that are more representative of future use cases~\cite{fiebrink2011human}. However, if it is an application intended to uncover machine learning concepts (\eg \cite{hitron2019can}) perhaps promoting more model-breaking examples~\cite{wallace2019trick} would be more appropriate; though in the context of a teachable interface this could lead to users training the model with less authentic data to simply improve its performance~\cite{zimmermann2019youth}.

This work has several limitations listed below: 

\textit{Task}: We explore machine teaching in a narrow context, that of a supervised 3-way image classification task. This allows us to dive deep in our analysis using a fine-grained scheme when coding participants examples informed from prior work on visual object understanding. However, it also limits the generalizability of our findings. We attempt to overcome this by connecting our results with that of prior work when possible. Three, the smallest number for multiclass classification, was selected to minimize challenges in finding different object instances within a category in a real-world environment as well as the task completion time (already 40 minutes long).

\textit{Study}: While teachable object recognizers are real-world applications~\cite{lee2019revisiting}, they are typically intended for blind users. Thus, the sighted participants may lack motivation in this study. We attempt to compensate for this lack of incentives with a performance-based payment scheme~\cite{ho2015incentivizing} creating the impression that we have a `secret' test to distinguish models that are more `robust'; though on our end this is merely a naive quality examination. By doing so, combined with the fact that the testbed shows only the predicted labels but no confidence scores in testing, we might have limited participants' criteria for model evaluation~\cite{fiebrink2011human} to just correctess. 

\textit{Analysis}: Through crowdsourcing we were able to quickly recruit a large participant pool and collect data outside a lab in the users' environment. However, this limited our control over the object instances that participants could use as well as the opportunity to create our own evaluation set for comparing the performance of the models against the same data. \\
To allow some time before testing for the photos to be received on our server and the models to be trained on our GPUs,  participants were asked to review their training photos and select 10 out of 30, 5 out of 10, and 1 out of 5. We are still analyzing these data while considering more fine-grained variation and inconsistency attributes.

\section{Conclusion and Future Work}
We have presented a crowdsourcing study, where MTurkers choose three objects in their environment and iteratively train a model to distinguish between them in real-time using the camera on their mobile phones. By doing so, we were able to explore, with a large participant pool ($N=100$), an instance of a machine teaching problem with a task where many non-experts can serve as the oracle.  Our findings and insights can contribute to the ongoing discussion on how non-experts conceptualize, experience, and reflect on their engagement with machine teaching. To allow for study replicability and future comparisons, we have provided a detailed description of our testbed, its framing within the machine teaching problem space from Zhu \etal~\cite{zhu2018overview}, and the list of questions and prompts used in the study.

Our results are based on a fine-grained analysis of the participants' examples contextualized by their responses, background, and model performance. We discuss how they can guide the design of future teachable interfaces to anticipate users tendencies, misconceptions, and assumptions. Given our research group's interest in teachable interfaces for accessibility~\cite{kacorri2017teachable}, our next step will be to explore whether these insights and data from sighted participants could be leveraged for the design of effective teachable object recognizers for blind users. Our rationale is that insights from this study can perhaps enable us to decouple non-experts misconceptions from challenges in camera manipulations among blind users~\cite{lee2019revisiting}.

\section{Acknowledgments}
The authors would like to thank the anonymous reviewers for their insightful comments on an earlier draft of this paper. This work is supported by NSF (\#1816380). Kyungjun Lee is supported by NIDILRR (\#90REGE0008).

\balance{}

\bibliographystyle{SIGCHI-Reference-Format}
\bibliography{bibliography}

\end{document}